\documentclass{amsart}
\usepackage{amssymb}
\title{The Volterra system and topology of the isospectral variety
of zero-diagonal Jacobi matrices}
\author{Alexei V. Penskoi}
\address{Independent University of Moscow,
Bolshoy Vlasyevskiy per. 11,
119002 Moscow Russia \& 
Bauman Moscow State Technical University, Moscow, Russia.}
\email{penskoi@mccme.ru}
\date{}
\newtheorem{Proposition}{Proposition}
\DeclareMathOperator{\diag}{diag}
\DeclareMathOperator{\tr}{tr}
\begin{document}
\maketitle

Let us consider a symplectic manifold $(X^{2n},\omega)$ and an integrable
system with Hamiltonian $H$ and involutive integrals
$F_1=H, F_2,\dots,F_n.$ Let $X_F\subset X$ be a submanifold defined
by equations $F_1=c_1,\dots,F_n=c_n.$ This submanifold is called
a level surface of integrals.
The well-known Liouville-Arnold theorem~\cite{Arnold}
says that if $X_F$ is compact and connected, then it is a torus.
This makes investigating the topology of $X_F$ trivial. However, it turns out that in
some important examples of integrable systems such
a submanifold is compact but its topology is quite complicated.
This is due to the fact that in these examples
$X$ has points where
either $H$ is singular or $\omega$ is singular or degenerate,
and the Liouville-Arnold theorem does not apply.
However, in some of these situations 
it is possible to define the corresponding flow on the whole
submanifold $X_F$ and successfully use it to investigate the  topology of $X_F.$

The first example of this kind was described by C.~Tomei in
his paper~\cite{T},
where the topology of the isospectral variety of Jacobi matrices 
(i.e. real three-diagonal symmetric matrices) is investigated. 
This is a level surface of integrals
of the (open) Toda lattice. Using the Toda flow Tomei computed
the Euler characteristic of this variety.
Two years later D.~Fried~\cite{F} discovered that the stable and unstable
stratifications defined by the Toda flow are cell complexes. 
Using this fact he found the cohomology ring of
this isospectral variety by direct calculation.

Our goal is to investigate topology of the isospectral variety of real zero-diagonal
Jacobi matrices. Let us consider the variety $M_k$ of all $(k\times k)$-matrices
$L$ of the form $L_{ij}=c_i\delta_{i+1,j}+c_{i-1}\delta_{i-1,j}$
with fixed spectrum. 

\begin{Proposition}\label{spectrum}
a) The eigenvalues of such a matrix
are of the form $0$ (if $k$ is odd), 
$\pm\lambda_1,\pm\lambda_2,\dots$
b) If all eigenvalues are distinct
then $M_k$ is a compact smooth manifold and its topology
does not depend on the eigenvalues.
\end{Proposition}

\noindent{\bf Proof} of a) is clear.
The proof of b) is analogous to Tomei's
proof of the analogous statement for the isospectral variety
of Jacobi matrices~\cite{T}.

The manifold $M_k$ is a level surface of integrals
of the (open) Volterra system
$\dot{c}_i=\frac{1}{2}c_i(c^2_{i+1}-c^2_{i-1}),$
where $c_0=c_{k}=0.$
Usually this system is written in terms of the variables  $u_i=c_i^2.$
It is well-known that the Volterra system can be written in the Lax form
$\dot{L}=[L,A(L)].$ It follows from this Lax form
that the Volterra flow preserves the spectrum of $L,$
i.e. the Volterra flow is a flow on $M_k.$ 
It is well known~\cite{M} that the Volterra system is
integrable.
In the even ($k=2l$) case $M_k$ has $2^l$ isomorphic
connected components. Moreover, each
component is isomorphic to the isospectral variety of Jacobi matrices
studied by Tomei and Fried, see e.g.~\cite{D,V}.
The odd case ($k=2l+1$) is completely different, 
$M_k$ is connected. It is not clear now how to compute the homology groups.
The stable and unstable stratifications are not cell
complexes in this case, and it is not possible to apply a direct approach
similar to Fried's~\cite{F}. We should also say that
in 1992 A.~Bloch, R.~Brockett and T.~Ratiu~\cite{BRR}
used the idea of the double bracket representation
to show that  the Toda flow is a gradient flow. 
It was shown in 1998~\cite{P} that the Volterra flow is also
a gradient flow. This leads us  to the natural idea
of finding the homology groups using the Morse complex. 
Unfortunately we cannot use the Morse complex for
computing the homology groups because the stable and unstable 
stratifications are not transversal to each other. 
In the present paper we compute the Euler characteristic 
of $M_k$ using the Volterra flow. The approach is similar to
Tomei's~\cite{T}.

Let $K=\frac{1}{4}\diag(1,2,3,\dots)$ and $f(L)=\tr KL^2.$
It has been shown~\cite{P} that the (negative) gradient flow
of $f(L)$ with respect to some metric coincides with
the Volterra flow. Let us choose $\lambda_i$
(see Proposition~\ref{spectrum}) in such a way that
all $\lambda_i>0$ and $\lambda_i>\lambda_{i+1}.$

\begin{Proposition}\label{critical}
a) The critical points of $f(L)$ on $M_{2l+1}$ are in one-to-one
correspondence with triples $[j,s,\pi]$
consisting of a number $j$ with $0\leqslant j\leqslant l,$ an $l$-tuple
$s=(s_1,\dots,s_l)$ with each $s_i$ equal to $0$ or $1,$
and a permutation $\pi\in S_l.$ b) The index of a critical point
corresponding to $[j,s,\pi]$ is equal to the number of
indexes $0\leqslant i\leqslant l-1$ with $i\ne j,j-1$ and
$\pi(i)<\pi(i+1),$ plus $1$ if $j\ne l.$
\end{Proposition}

\noindent{\bf Proof.} The critical points are equilibrium points
of the Volterra flow. Using this fact and the explicit polynomial
equations of $M_k$ one can prove that the critical points are
exactly those points with exactly $l$ of the $2l$ values $c_1,\dots,c_{2l}$ 
equal to zero and the additional property that $c_i\ne0$
implies $c_{i-1}=c_{i+1}=0.$
Looking at spectrum of the corresponding matrices one obtains the description
in the statement of the proposition. 
This proves a).
The formula b) for the index may be 
obtained by studying the Hessian
of $f(L).$ 

\begin{Proposition}
a) The Euler characteristic of $M_{2l+1}$ is equal to
$\chi(M_{2l+1})=2^{2l+2}(2^{l+2}-1)\frac{B_{l+2}}{l+2},$
where $B_{l+2}$ is a Bernoulli number.
b) If we define $\chi(M_1)$ to be $0,$ then the exponential generating function
is equal to $-\tanh^2(2z),$ i.e., 
$-\tanh^2(2z)=\sum_{l\geqslant0}\chi(M_{2l+1})\frac{z^l}{l!}.$
\end{Proposition}

\noindent{\bf Proof.} Recall that an interval
$[i,i+1]$ such that $\pi(i)<\pi(i+1)$ is called an ascent.
The number of permutations of $n$ elements with $k$
ascents is called the Euler number $\genfrac{<}{>}{0pt}{}{n}{k},$
see~\cite{K}.
Denote the number of ascents in $\pi$ by $p(\pi).$
Let $\psi(n)=\sum_{m=0}^n(-1)^m\genfrac{<}{>}{0pt}{}{n}{k},$
then formula (7.56) in~\cite{K} implies that
$1+\tanh z=\sum_{n\geqslant0}\psi(n)\frac{z^n}{n!}.$
Proposition~\ref{critical} implies that the Euler
characteristic $\chi(M_{2l+1})$ is equal to
$2^l\sum\limits_{j=1}^{l-1}\binom{l}{j}%
\sum_{\pi_1\in S_j,\pi_2\in S_{l-j}}(-1)^{p(\pi_1)+p(\pi_2)+1}=%
-2^l\sum\limits_{j=1}^{l-1}\binom{l}{j}%
(\sum_{\pi_1\in S_j}(-1)^{p(\pi_1)})
(\sum_{\pi_2\in S_{l-j}}(-1)^{p(\pi_2)})=%
-2^l\sum_{j=1}^{l-1}\binom{l}{j}\psi(j)\psi(l-j).$ 
This implies b).
Using the expansion of $\tanh z$ and the formula $\tanh'z=1-\tanh^2z,$
one obtains the formula from a).

The author is indebted to A.~P.~Veselov for attaching his
attention to this problem, and to O.~Cornea and A.~Medvedovsky for 
fruitful discussions.

\end{document}